\documentclass[prb,twocolumn,showpacs,amsmath,amssymb]{revtex4}

\usepackage{graphicx}
\usepackage{dcolumn}
\usepackage{bm}
\usepackage{color}

\def\tmo{TbMnO$_3$}
\def\ph#1{\phantom{#1}}
\def\bmult#1\emult{\begin{multline}#1\end{multline}}
\def\be{\begin{equation}}
\def\ee{\end{equation}}

\begin{document}

\title{Dependence of electronic polarization on
octahedral rotations in \tmo\ from first principles}

\author{Andrei Malashevich}
\email{andreim@physics.rutgers.edu}
\author{David Vanderbilt}
\affiliation{
Department of Physics \& Astronomy, Rutgers University,
Piscataway, NJ 08854-8019, USA
}

\date{\today}

\begin{abstract}
The electronic contribution to the
magnetically induced polarization in orthorhombic \tmo\ is studied
from first principles.  We compare the cases in which the spin cycloid,
which induces the electric polarization via the spin-orbit interaction,
is in either the $b$--$c$ or $a$--$b$ plane.  We find that the
electronic contribution is negligible in the first case, but much
larger, and comparable to the lattice-mediated contribution, in the
second case.  However, we show that this behavior is an artifact
of the particular pattern of octahedral rotations characterizing the
structurally relaxed $Pbnm$ crystal structure.  To do so, we
explore how the electronic contribution varies for a structural
model of rigidly rotated MnO$_6$ octahedra, and demonstrate that
it can vary over a wide range, comparable with the lattice-mediated
contribution, for both $b$--$c$ and $a$--$b$ spirals.  We introduce a
phenomenological model that is capable of describing this behavior
in terms of sums of symmetry-constrained contributions arising
from the displacements of oxygen atoms from the centers of the
Mn--Mn bonds.
\end{abstract}


\maketitle

\section{Introduction}
\label{sec:intro}

Multiferroic materials have been the subject of much
excitement because they exhibit many interesting properties and
phenomena.\cite{fiebig,kimura:2007,khomskii,eerenstein,cheong-mostovoy,tokura} 
There are two basic scenarios for the coexistence
of two order parameters in a single phase.  Either the two
instabilities can both be present independently, or one can induce
the other via some coupling mechanism.  We are concerned here
with magnetoelectric (ME) materials of the latter type, in which
the primary instability is magnetic, and the resulting magnetic
order induces an electric polarization.  Such magnetically-induced
(improper) ferroelectrics can be expected to display strong ME
couplings, e.g., a strong dependence of the electric polarization
on applied magnetic field, or of the magnetization on the
applied electric field.\cite{tokura}
They could be extremely useful in many
technological applications, but most of the materials discovered
to date either have too small of a ME coupling, or only operate
at impractically low temperatures.  Thus, it is essential to
understand their coupling mechanisms more fully in order to design
new materials with enhanced ME effects that can operate at higher
temperatures.

Among the best studied of the magnetically induced ferroelectrics
are orthorhombic rare-earth manganites, $o$--$R$MnO$_3$.
\cite{Lorenz:2007,picozzi:2007,Ren:2009,kimura,Goto:2004,cheong-mostovoy,
Xiang:2008,malash,malash_epjb}
Intensive experimental and theoretical studies have clarified many questions
regarding the origin of ferroelectricity in these compounds.
In HoMnO$_3$ and YMnO$_3$, the collinear $E$-type antiferromagnetic (AFM)
spin order induces a polarization through the exchange
striction mechanism.\cite{Lorenz:2007,picozzi:2007,Ren:2009}
In \tmo\ and DyMnO$_3$, in contrast, the polarization appears
with the onset of spiral magnetic order\cite{kimura,Goto:2004}
as a consequence of the spin-orbit interaction.\cite{cheong-mostovoy}

On a microscopic level, the appearance of the polarization
can result either from a change in electron charge density
that would occur even with ionic coordinates clamped (the purely
electronic contribution), or from displacements of
the ions away from their centrosymmetric positions as a result
of magnetically induced forces (lattice contributions).
In any theoretical analysis of such materials, it
is important to distinguish between these two contributions and to
calculate them separately in order to understand which microscopic
mechanisms are responsible for the appearance of the polarization.
\cite{picozzi:2007,Xiang:2008,malash,malash_epjb,Ren:2009}
Regarding the $o$--$R$MnO$_3$ materials having the cycloidal
spin structure, Xiang {\it et al.}\cite{Xiang:2008} and
we\cite{Xiang:2008,malash} have demonstrated that in \tmo\ the electronic
contribution to the polarization is much smaller than the lattice
contribution when the spin spiral is lying in the $b$--$c$
plane. For this reason, our previous work\cite{malash,malash_epjb}
focused on a detailed analysis of the lattice contribution,
while the electronic contribution was not studied carefully.
The mode decomposition of the lattice contribution\cite{malash}
and its dependence on the spin spiral wave vector\cite{malash_epjb}
revealed that the next-nearest-neighbor spin interactions are
important not only for the formation of the spin spiral itself,
but also for the induced polarization.

Although in this particular case the lattice contribution is
dominant, it is not clear how general this result is. Picozzi
{\it et al.}\cite{picozzi:2007} showed that for orthorhombic HoMnO$_3$,
in which the polarization is induced by collinear E-type AFM order,
the electronic
contribution to the polarization is of the same order as the
lattice contribution. Although the mechanism of polarization
induction in HoMnO$_3$ is different from that in \tmo, there is
no {\it a priori} reason why the electronic contribution should
be negligible in \tmo.  In fact, the first-principles study by
Xiang {\it et al.}\cite{Xiang:2008} found that if the spin spiral lies in the
$a$--$b$ plane, the electronic contribution to the polarization
is of the same order of magnitude as the lattice contribution.

In this work, we study the polarization induced
by the $a$--$b$-plane spin spiral and show that this case
differs significantly from the case of the $b$--$c$-plane spiral.
We focus mainly on the electronic contribution, analyzing it
carefully for both cases by considering how it varies for a
structural model of rigidly rotated MnO$_6$ octahedra.
We confirm the finding of Xiang {\it et al.}\cite{Xiang:2008}
that the purely electronic contribution to the polarization
is not negligible for the case of the $a$--$b$ spiral.  Indeed,
we find it to be quite sensitive to the choice of the calculation
parameters, as well as on the octahedral rotation angles.
Even for the case of the $b$--$c$ spiral, we find that the electronic
contribution can be quite significant if the octahedral rotation
angles are varied away from the equilibrium values.

We also construct a phenomenological model based on a symmetry
analysis of the spin-orbit induced electronic dipoles associated
with centrosymmetry-compatible oxygen displacements
relative to the centers of the Mn--Mn bonds,
finding that it is essential to take the Jahn-Teller orbital ordering
into account from the outset.
This model shows that $b$--$c$ and $a$--$b$ spin spirals need not be similar
in terms of how the polarization is induced.
Our work implies that the electronic contribution to the
polarization is generically expected to be much larger than was
found for the specific case of the relaxed $b$--$c$ spiral state
in TbMnO$_3$, emphasizing the importance of considering both
electronic and lattice mechanisms in any future theoretical studies
of this class of materials.

The rest of the paper is organized as follows.
In Sec.~\ref{sec:spirals} we compare the electric polarization
computed for fully relaxed \tmo\ with the spin spiral lying
in the $b$--$c$ and $a$--$b$ planes for several values of the on-site
Coulomb energy $U$ in the LDA+U framework. In Sec.~\ref{sec:modeling}
we focus on the study of the purely electronic contribution
to the polarization in the context of a structural model of
\tmo\ in which the Mn and O ions form rigid octahedra. The dependence
of the Berry-phase polarization on octahedral rotations is studied in 
Sec.~\ref{sec:str_model}, while in Sec.~\ref{sec:phenom} we develop
a symmetry-based phenomenological model in an attempt to explain the
observed results.  We discuss our findings in Sec.~\ref{sec:disc},
and give a brief summary in Sec.~\ref{sec:summary}.

\section{Spin spirals in the \lowercase{\emph{b-c}} and
\lowercase{\emph{a-b}} planes}
\label{sec:spirals}

A spiral (or, more precisely, ``cycloidal''\cite{explan:cycloid})
spin structure forms in \tmo\ in the $b$--$c$-plane
below $\sim$27\,K with the polarization lying along the
$\bf\hat{c}$ axis.\cite{kimura}
However, a sufficiently
strong magnetic field applied along the $\bf\hat{b}$ direction causes the
polarization to change its direction from $\bf\hat{c}$ to $\bf\hat{a}$
(``electric polarization flop''). It was suggested, and recently
confirmed,\cite{aliouane:2009} that this polarization flop results from
the change of the spin spiral from the $b$--$c$ to the $a$--$b$ plane
(``spin flop''); the polarization simply follows the spin spiral.
We shall refer to these two magnetic states as the
`$b$--$c$ spiral' and `$a$--$b$ spiral'. The former is
incommensurate with a wavevector $k_{\mathrm{s}}\simeq0.28$, while the
latter is commensurate with $k_{\mathrm{s}}=1/4$.

In this section we first review the main results of our previous
calculations\cite{malash,malash_epjb} of the polarization induced by
the $b$--$c$ spiral.
We then present our new calculations for the case of the
$a$--$b$ spiral, and compare these two cases.

We use a 60-atom supercell consisting of three $Pbnm$ unit cells,
corresponding to a spin-spiral wave vector of
$k_{\rm s}=1/3$, for both the $a$--$b$ and $b$--$c$ spirals.
Although $k_{\mathrm{s}}=1/4$ experimentally for the
$a$--$b$ spiral, the use of four unit cells would
be more computationally demanding, and the consistent use of
$k_{\rm s}=1/3$ facilitates comparisons between the two cases.
(An additional reason why $k_{\rm s}=1/3$ is more
convenient will be mentioned at the end of Sec.~\ref{sec:phenom}.)

Our electronic-structure calculations are carried out using a
projector augmented-wave\cite{Blochl,Kresse:1999} method implemented
in the VASP code package.\cite{kresse-vasp} Since the
local-density approximation (LDA) gives a metallic state for TbMnO$_3$,
we use on-site Coulomb corrections (LDA+U) in a rotationally invariant
formulation.\cite{Dudarev} The electric polarization
is computed using the Berry-phase method.\cite{King-Smith}

In our previous work on the $b$--$c$ spiral,\cite{malash,malash_epjb}
the structural relaxation was performed in the absence of
spin-orbit interaction (SOI), and we confirmed that
no polarization is induced by the magnetic order in this case.
In the presence of SOI, an electric polarization was found to appear.
We decomposed it into electronic and lattice contributions
by first keeping the ionic positions frozen at their centrosymmetric
values while computing $P$, and then by repeating the calculation after
allowing ions to relax.  We found the electronic and lattice contributions
to be $P_{\rm elec}=32$\,$\mu$C/m$^2$ and
$P_{\rm latt}=-499$\,$\mu$C/m$^2$ respectively. This result
demonstrated that the lattice mechanism dominates over the purely
electronic one for the $b$--$c$ spiral in TbMnO$_3$.

The polarization values quoted above were calculated using
$U=1$\,eV to match the experimental band gap.\cite{malash}
We also studied the effect of the choice of $U$ parameter (in a
reasonable range of values from 1\,eV to 4\,eV ) on the induced
polarization and coupling mechanism.\cite{malash_epjb}
We found that while
the absolute value of $P$ becomes somewhat smaller with larger $U$,
the qualitative mechanism of polarization induction remains the same.

In the case of the $a$--$b$ spiral, we have now performed similar
calculations of the polarization for the same
set of $U$ values from 1\,eV to 4\,eV.
Table~\ref{tab:pol_compare} shows the results for $U=1$\,eV and
$U=4$\,eV.  For these values we proceeded as in our previous
work, taking a reference crystal structure
that was fully relaxed in the absence of SOI, computing the
SOI-induced electric polarization $P^{\rm elec}$ and ionic forces,
and then using the latter, together with the computed Born
charges and force-constant matrix elements, to predict $P^{\rm latt}$.
(For intermediate $U$ we only computed $P^{\rm elec}$, and did so
in a simplified manner by using the reference crystal structure
that was relaxed at $U=1$\,eV, finding $P^{\rm elec}=691$ and 397\,$\mu$C/m$^2$
for $U=2$ and 3\,eV, respectively. These values are intermediate
between the values for $U=1$ and $U=4$\,eV as expected.)

\begin{table}
\caption{\label{tab:pol_compare}
Purely electronic, lattice, and total polarizations along
the $\bf\hat{c}$ and $\bf\hat{a}$ axes
for the $b$--$c$ and $a$--$b$ spirals respectively.
Results at $U_{\rm Mn}=2$\,eV (and also using $U_{\rm Tb}=6$\,eV) from
Ref.~\onlinecite{Xiang:2008} are shown for comparison.
}
\begin{ruledtabular}
\begin{tabular}{lrrrr}
spiral & $U_{\rm Mn}$ & $P^{\rm elec}$~~ & $P^{\rm latt}$~~
                                                         & $P^{\rm tot}$~~ \\
       & (eV)         & ($\mu$C/m$^2$)  & ($\mu$C/m$^2$)   & ($\mu$C/m$^2$)  \\
\hline
$b$--$c$ & 1\ph{$^*$} &     32~~        & $-$499~~         & $-$467~~        \\
         & 4\ph{$^*$} &  $-$14~~        & $-$204~~         & $-$218~~        \\
         & 2\footnotemark[1] & 1\footnotemark[1]\, &
           $-$425\footnotemark[1]\, & $-$424\footnotemark[1]\!
\smallskip \\
%
$a$--$b$ & 1\ph{$^*$} &     1530~~      & $-$790~~         &    740~~        \\
         & 4\ph{$^*$} &      174~~      & $-$197~~         & $-$23~~         \\
         & 2\footnotemark[1] & 331\footnotemark[1]\, &
           $-$462\footnotemark[1]\, & $-$131\footnotemark[1]\, \\
\end{tabular}
\footnotetext[1] {From Ref.~\onlinecite{Xiang:2008}.}
\end{ruledtabular}
\end{table}

For comparison, we also show in Table~\ref{tab:pol_compare} the results
of similar calculations by Xiang {\it et al.},\cite{Xiang:2008}
who used $U=2$\,eV on the Mn sites.  However, their results are not
directly comparable with ours, as they also included Tb $f$ electrons,
with $U_{\rm Tb}=6$\,eV on the Tb sites.
In the $b$--$c$ spiral case, as we mentioned before,
the results do not depend strongly on the choice of $U$. Comparing in this case
the results of Xiang {\it et al.}\ with our results for $U=1$\,eV, we find
an agreement in that the purely electronic contribution is negligible,
and the total polarization values agree with each other within 10\%.
However, in the $a$--$b$ spiral case there is no such agreement, which
is perhaps not surprising in view of the very strong sensitivity
of the polarization to the value of $U$, as can be seen clearly
in the table.
It may also result in part from other factors, such as the different
treatment of $f$ electrons in the two calculations, or the fact that
they used a generalized-gradient (GGA) exchange-correlation while we used LDA.
Nevertheless, a point in common is that both calculations predict that the
electronic contribution is comparable or even larger than the
lattice one for the $a$--$b$ spiral. This leads us to conclude that the
dominance of the lattice contribution that was found earlier for the case
of $b$--$c$ spiral is not a general phenomenon, but was special to that case.

For the $b$--$c$ spiral, the theoretical values in
Table~\ref{tab:pol_compare} are in satisfactory
agreement with the value of $\sim$$-$600\,$\mu$C/m$^2$ found
experimentally.\cite{kimura}  However, for the case of the 
$a$--$b$ spiral, the comparison is more problematic.
Our computed polarization of 740\,$\mu$C/m$^2$
compares very poorly with the
experimental value of $\sim$$-$300\,$\mu$C/m$^2$ obtained by Yamasaki
{\it et al.}\ for the related Gd$_{0.7}$Tb$_{0.3}$MnO$_3$ system.\cite{Yamasaki:2008}
However, as we shall discuss in Sec.~\ref{sec:disc}, the polarization
depends sensitively on the octahedral tilting angles, which may
differ significantly for Gd$_{0.7}$Tb$_{0.3}$MnO$_3$.
Experiments on the $a$--$b$ spiral in
the TbMnO$_3$ system itself are somewhat ambiguous regarding both the
sign and the magnitude of the polarization.\cite{kimura}

\begin{figure}
\centering \includegraphics[width=3.5in]{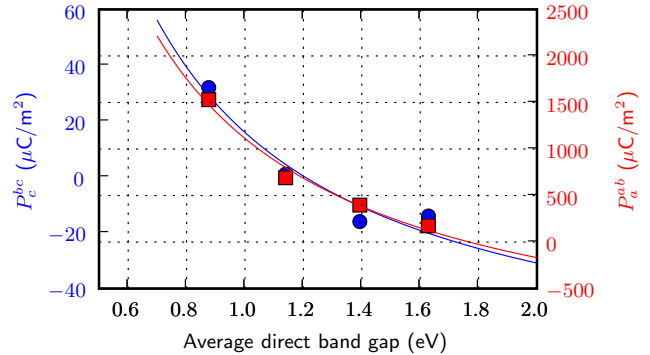}
\caption{(Color online.)
Dependence of electronic contribution to the polarization on
the average direct band gap
for the $b$--$c$ spiral (circles, scale at left) and $a$--$b$ spiral
(squares, scale at right) when varying $U$.
}
\label{fig:udep}
\end{figure}

Note that the magnitude of the electronic contribution in
the case of the $a$--$b$ spiral falls rapidly with increasing
$U$. Our calculations show that the band gap increases almost
linearly with $U$.  Fig.~\ref{fig:udep} shows the electronic
contribution to the polarization for both $a$--$b$ and $b$--$c$
spirals plotted versus the average direct band gap.  One can
see from the plot that the polarization is roughly inversely
proportional to the gap, up to a constant shift.
A heuristic rationalization of this behavior can be given
as follows.  If we consider the \emph{derivative} of the
polarization with respect to ionic displacements, which is the
Born effective charge, we know that this quantity can be expressed
within density-functional perturbation theory in a Kubo-Greenwood
form involving a sum over terms that are inversely proportional
to the differences of the eigenenergies of the unoccupied and
occupied states.\cite{Baroni_DFPT}  The largest contributions
are expected to come from the smallest energy denominators associated
with states near the valence and conduction edges, so the overall
sum should roughly scale inversely with the direct band gap.
The same applies to other derivatives of the polarization, such
as the dielectric susceptibility.  If the derivatives of $P$
have this behavior, it is not very surprising to find that the
polarization itself has a similar behavior.

In view of the results discussed above, the central question arises:
Why are the cases of the $a$--$b$ and $b$--$c$ spiral so different?
In the remainder of this paper, we attempt to shed some light on this question.
For this purpose, we limit ourselves to a discussion
of the purely electronic contribution to the polarization.
We shall discuss at some length the dependence of the electronic
polarization on atomic displacements, but only for displacement patterns
that preserve inversion symmetry, such as those resulting from octahedral
tilting in the $Pbnm$ crystal structure.

\section{Modeling of $P_{\rm elec}$}
\label{sec:modeling}

\subsection{Structural model with rigid MnO$_6$ octahedral rotations}
\label{sec:str_model}

To obtain a better understanding of the mechanism of the electronic
contribution to the polarization, we consider a simplified
structural model in which the crystal structure is composed of
rigid corner-linked MnO$_6$ octahedra, with the Tb ions
remaining at their high-symmetry (0,0,1/4) Wyckoff coordinates.
We then rotate the MnO$_6$ octahedra
and study how the polarization depends on the rotation angles.
All calculations within this model are done with $U=1$\,eV.
While the values of the polarization computed with this $U$ may not
be realistic for the case of the $a$--$b$ spiral, as discussed above,
at this point we are interested in understanding the origins and
behavior of the polarization, rather than making direct comparisons to
experiment. Also, we want to compare the results with previous
calculations,\cite{malash,malash_epjb} most of which were done at $U=1$\,eV.

Actually, before we even apply the rotations,
we must first apply a Jahn-Teller (JT) distortion.
Mn$^{3+}$ has a $d^4$ configuration in which the three majority-spin
$t_{2g}$ states are filled and the majority-spin $e_g$ levels are
half-filled.  The system is thus metallic in the absence of the JT
distortion; introducing it splits the
$e_g$ levels and opens a gap, driving the system insulating.
In our model, we take the JT distortion into account by pre-deforming the
MnO$_6$ octahedra such that the ratio of longest to intermediate (along
$c$) to shortest bonds lengths is 1.124\,:\,1.004\,:\,1, where these
ratios have been extracted from our earlier first-principles calculations
carried out with $U=1$\,eV.\cite{malash}
\begin{figure}
\centering \includegraphics[width=3.0in]{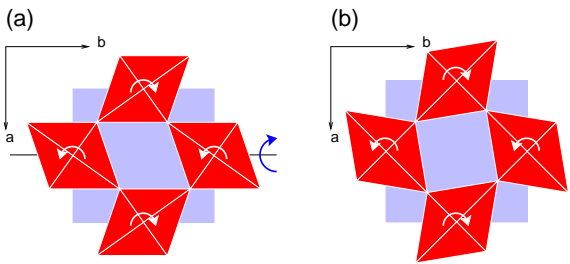}
\caption{(Color online.)
Two initial configurations considered in the model of rigid
MnO$_6$ octahedra.
Rotations about $\bf\hat{c}$ and $\bf\hat{b}$
axes are indicated by white and dark blue curved arrows respectively.
(a) Structure 1, which matches the physical $Pbnm$ structure fairly
closely.  (b) Structure 2, with a fictitious pattern of octahedral
rotations around the $\bf\hat{c}$ axis only.
}
\label{fig:octa}
\end{figure}

We then apply rotations to the octahedra.
In $Pbnm$ symmetry the rotations can be described
as $(a^{-}a^{-}b^{+})$ in the Glazer notation,\cite{glazer} meaning
that out-of-phase and in-phase alternating rotations occur around
$[110]$ and $[001]$ axes respectively in the
original cubic-perovskite Cartesian frame.
In the conventional frame \cite{kimura} used here,
these correspond to the $\bf\hat{b}$ and $\bf\hat{c}$ axes respectively, and the
spin-spiral wave vector propagates along $\bf\hat{b}$.

In general, a different order of application of rotations leads to different
final configurations, because rotations do not commute.
Therefore, when describing the \tmo\ system in terms
of MnO$_6$ octahedral rotations, one should carefully specify the meaning
of the rotations and their order.
For example, if we start with the ideal
perovskite configuration and induce the Jahn-Teller distortion, we can
arrive at several possible initial configurations as shown
in Fig.~\ref{fig:octa}. If we apply $(a^{-}a^{-}b^{+})$ rotations
to the configuration shown in Fig.~\ref{fig:octa}(b), regardless of the
order of rotations, the final structure will not have $Pbnm$ symmetry.
However, applying a rotation around $\bf\hat{b}$ followed by a rotation around
$\bf\hat{c}$ to the configuration shown in Fig.~\ref{fig:octa}(a), we
will preserve the $Pbnm$ symmetry.

Fitting the angles of rotation to the relaxed structure, we find the rotation
angles around $\bf\hat{b}$ and $\bf\hat{c}$ to be approximately $19.0^{\circ}$
and $11.6^{\circ}$ respectively.  We thus constrain the ratio between
these two angles to be 1.64, and treat the angle $\theta$ around the
$\bf\hat{b}$ axis as the independent variable.  We calculate the polarization
for both the $a$--$b$ and $b$--$c$ spirals for a range of rotation
angles 
$-15^{\circ}<\theta<20^{\circ}$. We also made several
calculations for the initial configuration shown in Fig.~\ref{fig:octa}(b),
where the octahedra were rotated only around $\bf\hat{c}$. To distinguish between
the two sets of calculations, we will refer to them as `Structure 1' and
`Structure 2' respectively.

\begin{figure}
\centering \includegraphics[width=3.0in]{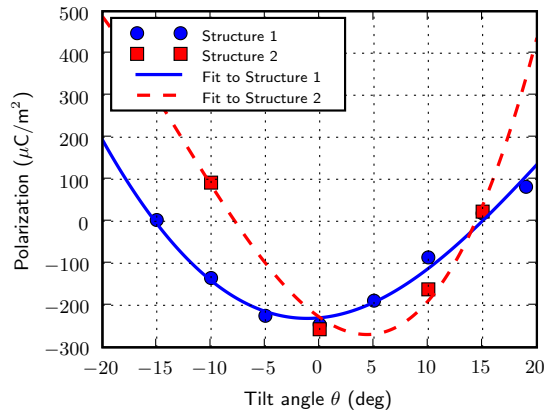}
\caption{(Color online)
For $b$--$c$ spiral, the dependence of the electronic contribution
to the polarization on $\theta$, the angle of rotation of the
MnO$_6$ octahedra about the $\bf\hat{b}$ or $\bf\hat{c}$ axis for
Structure 1 or 2 respectively.  (For the former, the rotation
angle around $\bf\hat{c}$ is $\theta/1.64$; see text for details.)
Symbols: first-principles calculations.
Curves: result of the fit to the phenomenological model of
Sec.~\ref{sec:phenom}.
}
\label{fig:fit_bc_n}
\end{figure}

\begin{figure}
\centering \includegraphics[width=3.0in]{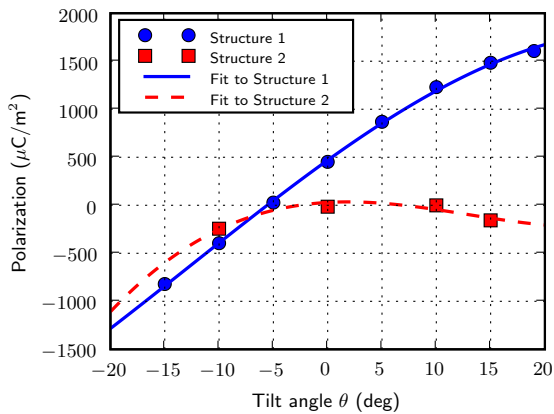}
\caption{(Color online)
For $a$--$b$ spiral, the dependence of the electronic contribution
to the polarization on $\theta$, the angle of rotation of the
MnO$_6$ octahedra about the $\bf\hat{b}$ or $\bf\hat{c}$ axis for
Structure 1 or 2 respectively.  (For the former, the rotation
angle around $\bf\hat{c}$ is $\theta/1.64$; see text for details.)
Symbols: first-principles calculations.
Curves: result of the fit to the phenomenological model of
Sec.~\ref{sec:phenom}.
}
\label{fig:fit_ab_n}
\end{figure}

The results of the calculations are
presented in Figs.~\ref{fig:fit_bc_n} and \ref{fig:fit_ab_n}. Recall
that all structures considered here have inversion symmetry,
so that the Berry-phase calculations give us the purely electronic
contribution to the polarization induced by the SOI.
These calculations reveal that even in the case of the $b$--$c$
spiral, the electronic contribution to the polarization spans a wide
range of values ($\sim$300\,$\mu$C/m$^2$), depending on the octahedral
rotations. This is yet another indication that this contribution is
negligible in the relaxed $b$--$c$ spiral structure only by coincidence.

\subsection{Phenomenological model}
\label{sec:phenom}

\begin{figure}[b]
\centering \includegraphics[width=3.5in]{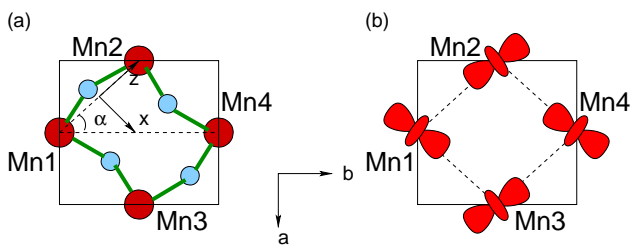}
\caption{
(a) Local (${x}$,${y}$,${z}$)
and global (${a}$,${b}$,${c}$) coordinate frames.
(b) Orbital ordering in TbMnO$_3$. The $d_{3x^2-r^2}/d_{3y^2-r^2}$ orbitals
are aligned along the longest Mn--O bonds. Orbital order is uniform along
the $\bf\hat{c}$ axis.
}
\label{fig:bonds}
\end{figure}

To find out whether the observed dependence of the polarization
on octahedral rotations can be explained within some relatively
simple model, we decided to analyze the possible contributions
coming from each nearest-neighbor Mn--O--Mn triplet.  Our notation
is as follows.  We use $\bf\hat{a}$, $\bf\hat{b}$, and $\bf\hat{c}$ for
the unit vectors in the global Cartesian frame of
Fig.~\ref{fig:octa}.  We also attach a local Cartesian frame to
each Mn--O--Mn triplet as illustrated in Fig.~\ref{fig:bonds}(a),
reserving ${\bf\hat{z}}={\bf\hat{e}}_{12}$ (the unit vector pointing
from Mn1 to Mn2), while $\bf\hat{x}$ and $\bf\hat{y}$ are chosen to
form a right-handed triad with $\bf\hat{z}$ such that
$\bf\hat{x}$ lies in the $a$--$b$ plane. The origin of this frame
is located in the middle of the Mn--Mn bond. The angle between
${\bf\hat{e}}_{12}$ and the spin-spiral wavevector direction $\bf\hat{b}$
is denoted by $\alpha$, so that $\cos\alpha={\bf\hat{z}}\cdot{\bf\hat{b}}$.
For the vertical bonds parallel to $\bf\hat{c}$, the local and
global Cartesian frames coincide and $\alpha=\pi/2$.

Our goal is to find dipole moments allowed by symmetry for each
Mn--O--Mn bond, viewed in isolation.
To find the total polarization, we need to transform these
dipole moments back into the global Cartesian frame,
average them over the spin-spiral
period, and sum the contributions from all Mn--O--Mn bonds in the
20-atom unit cell. One can show that there is no contribution to the
polarization coming from the vertical bonds by using the fact
that the magnetic moments on the Mn sites are collinear
for these bonds.  Therefore, we focus henceforth only on the bonds
lying in the $a$--$b$ plane.

We expand the dipole moment of each Mn--O--Mn triplet
in (i) bilinear products of spin components on the two Mn sites,
and (ii) powers of oxygen displacements related to the MnO$_6$
octahedral rotations.  We invoke symmetry to determine the appropriate
terms in this expansion, as follows.  Consider the configuration shown in
Fig.~\ref{fig:octa}(b). The Mn--O--Mn bonds have two mirror
symmetries, $M_x$ and $M_y$. Note that the bond does not have
inversion symmetry because the JT distortion leads to an orbital
ordering pattern, shown schematically in Fig.~\ref{fig:bonds}(b),
which breaks $M_z$.  We emphasize that it is essential to take this
JT distortion, and the associated orbital order, into account.  If
instead one attempts to use a JT-free perovskite structure as
the reference for such an expansion, one finds the reference
system to be metallic, so that the electric polarization cannot
even be defined.  Therefore, it is first necessary to establish
an orbitally-ordered insulating state, and only then expand
the electronic dipoles in lattice displacements away from that
state. Indeed, we initially attempted to derive a model built
on the erroneous assumption of inversion symmetry for Mn--O--Mn
bonds in a reference structure without JT distortions, but we
could not arrive at a satisfactory low-order expansion that could
simultaneously fit the data for both $a$--$b$ and $b$--$c$ spirals.

\begin{table}
\caption{\label{tab:mir_class}
Classification of several quantities by their behavior
under the mirror symmetries $M_x$ and $M_y$:
products of the spin components of two neighboring spins
${\bf S}_1$ and ${\bf S}_2$,
components of oxygen displacement vectors $\bf u$, and
components of the polarization vector $\bf P$.
}
\begin{ruledtabular}
\begin{tabular}{llrrrr}
 $M_x$ & $M_y$ &  &  &  &  \\
\hline
$+1$ & $+1$\ph{$\Big(\Big)$}  & $S_{1x}S_{2x}$, $S_{1y}S_{2y}$,
$S_{1z}S_{2z}$ & $u_z$ & $u_x^2$, $u_y^2$, $u_z^2$ & $P_z$ \\
$+1$ & $-1$\ph{$\Big(\Big)$}  & $S_{1y}S_{2z}$, $S_{1z}S_{2y}$ & $u_y$ & $u_yu_z$ & $P_y$ \\
$-1$ & $+1$\ph{$\Big(\Big)$}  & $S_{1x}S_{2z}$, $S_{1z}S_{2x}$ & $u_x$ & $u_xu_z$ & $P_x$ \\
$-1$ & $-1$\ph{$\Big(\Big)$}  & $S_{1x}S_{2y}$, $S_{1y}S_{2x}$ &  & $u_xu_y$ & \\
\end{tabular}
\end{ruledtabular}
\end{table}

We classify the products of spin components by their behavior under the
two mirror symmetries, as tabulated in Table~\ref{tab:mir_class}.
We can then systematically expand the polarization in powers of
displacements $(u_x,u_y,u_z)$ as
\bmult
\label{p_expan}
P_x=A^{(0)}_{xz}S_{1x}S_{2z}+A^{(0)}_{zx}S_{1z}S_{2x} \\
+[A^{(1)}_{xx}S_{1x}S_{2x}+A^{(1)}_{yy}S_{1y}S_{2y}
+A^{(1)}_{zz}S_{1z}S_{2z}]u_x \\
+[A^{(1)}_{xy}S_{1x}S_{2y}+A^{(1)}_{yx}S_{1y}S_{2x}]u_y \\
+[A^{(1)}_{xz}S_{1x}S_{2z}+A^{(1)}_{zx}S_{1z}S_{2x}]u_z+\dots
\emult
Here we show only the terms that appear at zero and
first order in $u$ because the expression rapidly becomes tedious
at higher order, but our analysis also includes all second-order terms.
Similar expressions
can be written for $P_y$ and $P_z$. Projecting these contributions
on the $\bf\hat{a}$, $\bf\hat{b}$, and $\bf\hat{c}$
axes and averaging over the spin-spiral period
and all Mn-Mn bonds in the unit cell, we arrive at
\bmult
\label{exp_bc}
P^{(b\mbox{-}c)}_c=\sin\phi\{C_0+C_xu_x+C_zu_z+C_{xx}u_x^2+C_{yy}u_y^2 \\
+C_{zz}u_z^2+C_{xz}u_xu_z\}
\emult
for the polarization in the case of the $b$--$c$ spiral, and
\bmult
\label{exp_ab}
P^{(a\mbox{-}b)}_a=\sin\phi\{A_0+A_xu_x+A_zu_z+A_{xx}u_x^2+A_{yy}u_y^2 \\
+A_{zz}u_z^2+A_{xz}u_xu_z\}
\emult
in the case of $a$--$b$ spiral.
The coefficients in Eqs.~(\ref{exp_bc}-\ref{exp_ab}) are just certain
linear combinations of those appearing in Eq.~(\ref{p_expan}) and the
corresponding equations for $P_y$ and $P_z$.

The resulting expressions in Eqs.~(\ref{exp_bc}-\ref{exp_ab}) for the
polarization may be viewed as simple Taylor expansions in the
oxygen displacements from the centers of the Mn--O--Mn bonds.
However, some terms are missing because they are forbidden by symmetry.
For example, terms linear in $u_y$ vanish after averaging along $\bf\hat{c}$
because the contribution from any bond in Fig.~\ref{fig:bonds}(a) is
canceled by the one from the bond above or below.  The symmetry
also implies that $P^{(b\mbox{-}c)}_a= P^{(b\mbox{-}c)}_b=
P^{(a\mbox{-}b)}_b= P^{(a\mbox{-}b)}_c=0$ after
averaging over the spin-spiral period.
The factor of $\sin\phi$ comes from averaging the products of
spin components over the spin-spiral period.
In this model we did not consider next-nearest-neighbor spin interactions,
which were shown to play an important role in the dependence of
$P_c$ on the magnitude of the wave vector in the $b$-$c$ spiral
case.\cite{malash,malash_epjb}
However, we argue that the contributions to the polarization
coming from these interactions will have essentially the same form
as Eqs.~(\ref{exp_bc}-\ref{exp_ab}), but with $\sin\phi$
replaced by $\sin(2\phi)$. In our particular case $\phi=\pi/3$,
so that $\sin\phi=\sin(2\phi)$ and the inclusion of the next-nearest-neighbor
interactions will only lead to a renormalization
of the coefficients in the expansions.

Eqs.~(\ref{exp_bc}-\ref{exp_ab}) have no coefficients in common,
showing that within this model there is no connection between
the polarization in the $b$--$c$ and $a$--$b$ spirals.
The results for each spiral case can be
fitted independently with seven parameters, whose fitted values are
given in Table~\ref{tab:newparams}.  In each of Figs.~\ref{fig:fit_bc_n}
and \ref{fig:fit_ab_n}, the resulting fits are shown as the solid
and dashed curves that refer, respectively, to Structures 1 and 2
of Fig.~\ref{fig:octa}(a) and Fig.~\ref{fig:octa}(b).
Eqs.~(\ref{exp_bc}-\ref{exp_ab}) clearly provide enough freedom to allow
a very good simultaneous fit to the results computed directly from first
principles for both structures.
If we go back and take the relaxed structures that were obtained
directly from the first-principles calculations (for $U$=1\,eV and
no SOI) and use the present model to evaluate the electronic
contribution to the polarization, we obtain
$P_c=1600\,\mu$C/m$^2$ and $P_a=31\,\mu$C/m$^2$
for the $b$--$c$ and $a$--$b$ spirals, to be compared with
values of 1530 and 32\,$\mu$C/m$^2$ computed directly from
first principles, respectively.

\begin{table}
\caption{\label{tab:newparams}
Fitted parameters $C_i$ and $A_i$ for the model
of Eqs.~(\ref{exp_bc}-\ref{exp_ab}) for $b$--$c$ and
$a$--$b$ spirals respectively.
Units are $\mu$C/m$^2$, $\mu$C/m$^2$\AA, and $\mu$C/m$^2$\AA$^2$
for terms of overall order 0, 1, and 2, respectively.
Column labels are subscripts $i$.
}
\begin{ruledtabular}
\begin{tabular}{crrrrrrr}
& $0$~ & $x$~ & $z$~ & $xx$~ & $yy$~ & $zz$~ & $xz$~ \\
\hline
$C_i$\ph{$\int^{\int}$} & $-$285 & $-$3466 & 211 & 1297 & 1143 & 3263 & $-$30576 \\
$A_i$\ph{$\int^{\int}$} & 92 & $-$8245 & 529 & 555 & $-$1403 & 361 & $-$33707 \\
\end{tabular}
\end{ruledtabular}
\end{table}

\section{Discussion}
\label{sec:disc}

Our Berry-phase calculations of the electronic contribution to the 
polarization for various model \tmo\ structures, 
in which rigid MnO$_6$ octahedra were rotated, show that the 
polarization in the $a$--$b$-spiral case behaves quite differently 
than for the $b$--$c$-spiral case. Not only is the range of values
different, but the qualitative dependence of $P_{\rm elec}$
on rotation angles is dissimilar. For the $b$--$c$ spiral,
the polarization shows a parabolic dependence on the rotation 
angles, while for the $a$--$b$ case the dependence is almost linear 
over a wide range of rotation angles. The phenomenological model
considered above suggests that, at least from the point of view
of symmetry, there is no relation between the $b$--$c$ spiral
and $a$--$b$ spirals, so that the observed differences should not
be very surprising.

Considering that the octahedral rotation angles and oxygen
displacements become quite large, we obtain quite good fits
of the dependence of the polarization on rotation angles from
our expansion of the dipoles on oxygen displacements away from
the Mn-Mn bond centers.  The observed behavior results from the
fact that the coefficients in front of $u_x$ and $u_z$ are much
larger in the $a$--$b$ spiral, while the quadratic coefficients
in front of $u_{xx}$ and $u_{zz}$ are much larger in the $b$--$c$
spiral (see Table~\ref{tab:newparams}).
It still remains to understand how the most important coefficients
in the expansion get their values based on some microscopic model
of bond hybridization.

If we compare our computed polarization for the relaxed TbMnO$_3$ structure 
in the $a$--$b$ spiral case (740\,$\mu$C/m$^2$, see Table~\ref{tab:pol_compare})
to the experimental value for Gd$_{0.7}$Tb$_{0.3}$MnO$_3$ in Ref.~\onlinecite{Yamasaki:2008}
($\sim$$-$300\,$\mu$C/m$^2$), we find very poor agreement.
However, since Gd has a larger radius than Tb, the MnO$_6$ octahedra will
be less tilted, reducing the electronic contribution to the polarization
(see Strucure 1 in Fig.~\ref{fig:fit_ab_n}).
This effect may help explain the observed difference.
However, it should also be kept in mind that
the strong sensitivity of the polarization to the choice of $U$ in the
case of the $a$--$b$ spiral means that any prediction
of the polarization made within the LDA+U framework will have a much
larger uncertainty than for the $b$--$c$ spiral case.
The use of linear-response techniques to compute the effective parameters
for the LDA+U method\cite{Matteo} could thus be appropriate here.
However, the very use of the LDA+U method itself may be questionable,
and it may be worth exploring the suitability of other methods,
such as GW quasiparticle \cite{gw} or dynamical mean-field theory
\cite{dmft} approaches, for computing the polarization in this case.

We have seen that the octahedral rotations can significantly change
the polarization.  Although  we have focused here only on the
electronic contribution, it appears likely
that the lattice contribution will also depend strongly on rotation
angles. Such a calculation of the lattice contribution
is problematic, however, because one wants to consider the
SOI-induced symmetry-breaking distortions away from a reference structure
that is not itself an equilibrium structure in the absense of SOI.
In principle it may be possible to compute these using an approach
similar to that in Ref.~\onlinecite{malash}.  That is, one would
compute the force-constant matrix and dynamical charges
(in the absence of SOI) and the SOI-induced forces for a given
configuration of octahedral rotations, and use these to predict
the induced amplitudes of the infra-red--active phonon modes and
the resulting lattice contribution to the polarization.  We have not
pursued such an approach here, as it would take us beyond the intended
scope of the present work.

\section{Summary}
\label{sec:summary}

We have used first-principles methods to compute the
electronic and lattice contributions to the spin-orbit induced
electric polarization in the cycloidal-spin compound \tmo\ with
the spin spiral in the $b$--$c$ and $a$--$b$ planes.
In the latter case we find that the electronic contribution
is of the same order of magnitude as the lattice contribution,
in strong contrast to previous studies of the $b$--$c$ case.

We have studied the electronic contribution to the polarization
in detail by considering a structural model based on rigid rotations
of MnO$_6$ octahedra.  We have shown that the electronic
contribution to the polarization can change significantly with
rotation angle even in the case of the $b$--$c$ spiral,
thus demonstrating that our previous neglect of this contribution
was justified only because of an accidental property of the
relaxed $Pbnm$ structure.
We have introduced a phenomenological model that expands the electronic
contribution to the polarization up to second order in the oxygen
displacements from the Mn--Mn midbond positions, and
have shown that it can explain the quite different behavior of the
polarization in the $b$--$c$ and $a$--$b$ spiral cases.


$\phantom{a}$

\acknowledgments
This work was supported by NSF Grant DMR-0549198.

\vfill

\bibliography{tmo}

\end{document}